\documentclass[aps,prd,twocolumn,showpacs,floatfix,nofootinbib,superscriptaddress]{revtex4-1}
\usepackage{amssymb}
\usepackage{graphicx}
\usepackage{xurl}
\usepackage{hyperref}
\usepackage{color}
\usepackage{mathtools}
\usepackage{times}
\usepackage{bm}
\usepackage{xcolor}
\usepackage{soul}
\usepackage{comment}

\usepackage[normalem]{ulem}

\newcommand{\up}[1]{{\rm #1}}

\newcommand{\Dquad}{\qquad\qquad}

\newcommand{\nnn}{\nonumber \\}

\newcommand{\beeq}{\begin{equation}}
	\newcommand{\eneq}{\end{equation}}
\newcommand{\bear}{\begin{eqnarray}}
	\newcommand{\enar}{\end{eqnarray}}

\newcommand{\rbar}{\bar r}       

\newcommand{\HH}{\mathcal{H}}   

\newcommand{\OO}{\mathcal{O}}

\newcommand{\al}{\alpha}
\newcommand{\be}{\beta}
\newcommand{\ga}{\gamma}
\newcommand{\de}{\delta}

\newcommand{\para}{\parallel}
\newcommand{\pa}{\partial}

\newcommand{\bobs}{\bar{\rm o}}

\newcommand{\RR}{\mathcal{R}}  

\begin{document}

\title{Infrared Sensitivity of Cosmological Probes In The Presence of Axion Field
Fluctuations}

\author{Matteo Magi}
\email{matteo.magi@uzh.ch}
\affiliation{Department of Astrophysics,
University of Zurich, Winterthurerstrasse 190,
CH-8057, Zurich, Switzerland}
\author{Robert Brandenberger}
\email{rhb@physics.mcgill.ca}
\affiliation{Department of Physics, McGill University, Montréal, QC, H3A 2T8, Canada}
\author{Jaiyul Yoo}
\email{jyoo@physik.uzh.ch}
\affiliation{Department of Astrophysics,
University of Zurich, Winterthurerstrasse 190,
CH-8057, Zurich, Switzerland}
\affiliation{Physics Institute, University of Zurich,
Winterthurerstrasse 190, CH-8057, Zurich, Switzerland}

\date{November 27, 2024}

\begin{abstract}

We study the effects of long wavelength entropy fluctuations on cosmological probes such as galaxy clustering, luminosity distance, and CMB temperature anisotropies. Specifically, we consider fluctuations of a massless spectator scalar field set up in the early universe, which later acquires mass during the radiation-dominated era. We find that there are non-vanishing effects on observables, and the amplitude of these effects peaks for observables set up at the time of equal matter and radiation, and decreases as~$\eta^{-2}$ where~$\eta$ is the conformal time. Hence, the back-reaction effects are important for CMB anisotropies, but their impact on late-time observables is suppressed. In particular, the back-reaction effects are unable to explain the Hubble tension while they might alleviate the cosmic dipole tension. In contrast to a lot of the previous work on back-reaction, we work in position rather than momentum space.

\end{abstract}

\maketitle

    \section{INTRODUCTION}
 
There has been a long-standing interest in the effects of long wavelength fluctuations on cosmological observables. One of the motivations for this interest is the work of Polyakov \cite{Polyakov} which indicates that infrared fluctuation modes induced by matter may destabilize de Sitter space and lead to a relaxation of an effective cosmological constant (see also \cite{other}). Later, perturbative calculations \cite{Woodard} showed that infrared gravitons can have the same effect. With an entirely different motivation, the back-reaction of long wavelength (super-Hubble wavelength) scalar metric fluctuations in inflationary cosmology was considered \cite{Abramo} indicating that these modes lead to a large and secularly growing effect on the background geometry which could \cite{RHB-Paris} lead to a dynamical relaxation of the cosmological constant.

The calculations of \cite{Woodard} and \cite{Abramo} involve computing the effects of quadratic combinations of linear fluctuations on the background geometry. However, in \cite{Unruh} an important concern was raised, namely whether a physical observer could actually detect the effect. It was then shown \cite{Ghazal1, AW} that in the case of purely adiabatic fluctuations, the effects computed in \cite{Abramo} correspond to a second order reparametrization of the time coordinate and that it cannot be measured by a physical clock.\footnote{This conclusion parallels the results of a number of authors (see e.g.  \cite{Weinberg,  Senatore}) which demonstrate that the back-reaction effects of infrared modes decouple in the case of adiabatic fluctuations.} On the other hand, in the presence of entropy fluctuations, infrared modes can have a measurable effect on the background \cite{Ghazal2}. In particular, the back-reaction effect of scalar metric fluctuations leads to a decrease in the average value of the measured Hubble expansion rate, where averaging is over a surface of constant entropy field \cite{Leila, Comeau}.

To make contact with actual observations, it is crucial to ask whether cosmological observables measured by tracing back along the observer’s past light cone are affected by long wavelength fluctuations. Averages of physical observables taken over fixed space-like surfaces are less closely connected with actual observations. We will be specifically interested in the effects of long wavelength fluctuations on the luminosity distance relationship, on galaxy clustering statistics, and on cosmic microwave background (CMB) temperature anisotropies.

The effects of metric fluctuations on the luminosity distance relationship were first studied in \cite{Sasaki} and then in more detail in \cite{LF}. Effects on galaxy clustering were then worked out in \cite{GC}, and consequences for CMB anisotropies in \cite{CMB}. Light-cone averaging was discussed in \cite{LC}. When applied to a cosmological model with an approximately scale-invariant spectrum of curvature fluctuations like would emerge from inflation \cite{ChibMukh}, for certain bouncing cosmologies \cite{Fabio} and for emergent cosmologies like String Gas Cosmology \cite{SGC, matrix}, there is at first sight the possibility of effects on observables which are infrared divergent \cite{IRproblem}.\footnote{See e.g. Appendix B of \cite{Bmode} and \cite{TCCrev} for assessments of the status of these cosmological models.} However, in the case of adiabatic fluctuations on super-Hubble scales, it was shown \cite{noIR} that the physical observables constructed assuming general relativity do not suffer from this problem.

In a previous work \cite{MY}, two of us have studied conditions under which observables such as the luminosity distance-redshift relation and the spectrum of CMB anisotropies are not affected by fluctuations with wavelength much larger than the characteristic scale of the probe. It was found that, independently of the specific form of the gravitational field equations, fluctuations need to be purely adiabatic on large scales.

On the other hand, non-adiabatic fluctuations are rather generic when matter consists of several fluids which are only weakly interacting, like e.g. in the present universe where dark matter and radiation are only very weakly coupled. Hence, it is interesting to study which of the conditions derived in \cite{MY} are broken and which observations will be affected.\footnote{Note that if the observer is not comoving with respect to the dominant matter, then we are effectively in the situation of having an entropy perturbation, the entropy field being the clock field. See \cite{Comeau} for a detailed discussion of this point.}

Specifically, we will consider a model in which matter fields consist of radiation and an axion field (which can be the dark matter) with initial conditions chosen early in the radiation phase such that the axion field is fluctuating relative to the constant density surface (which is the constant radiation temperature surface). Such initial conditions are generic if we consider an axion field (like the QCD axion \cite{QCDaxion}) for which the potential is generated at a specific time (the time at which QCD instanton effects become effective).

We find that in the presence of entropy fluctuations there are non-vanishing back-reaction effects on cosmological observables. The amplitude of these effects peaks for observables at the time of equal matter and radiation, and decays as~$\eta^{-2}$ for later time observables. Hence, the effects on CMB anisotropies can be large, but the effects on late-time observables are suppressed.

We use natural units in which $c=\hbar=k_{\text{B}}=1$. For simplicity, we consider a spatially flat background geometry, and we denote by~$x$ the comoving spatial coordinates, and by~$\eta$ the conformal time.

        \section{GENERAL CONDITIONS FOR IR-INSENSITIVE COSMOLOGICAL PROBES}
\label{sec2}

We assume a linearly perturbed FLRW universe as our cosmological model. Considering only scalar perturbations, the metric in an arbitrary gauge is described by the line element\footnote{See e.g. \cite{MFB, RHBfluctsrev} for reviews of the theory of cosmological perturbations.}
    \bear\label{metric}
        ds^2&=&-a^2(1 + 2\al)d\eta^2-2a^2\pa_i\be\,d\eta dx^i
        \nnn&&
        +a^2\left[( 1+2\varphi)\de_{ij}+2\pa_i\pa_j\ga\right] dx^i dx^j~,  
    \enar
where~$a(\eta)$ is the scale factor as a function of conformal time~$\eta$,~$x^i$ are comoving coordinates, and~$\al$,~$\be$,~$\ga$, and~$\varphi$ are linear scalar perturbations depending on the spacetime position~$x^\mu$.

We further specify the contents of this universe as a perfect fluid consisting of the mixture of two perfect fluids: a dust fluid describing non-relativistic matter, and a relativistic fluid describing radiation. The energy-momentum tensor takes the form
   \beeq\label{energymomentum}
        \begin{split}
        -T^0{}_0=\rho=\rho_m+\rho_\ga \,,\qquad\quad  T^i{}_j=p\,\de^i_j= p_\ga\de^i_j\,,
        \\
        -T^0{}_i=\left(\bar p+\bar\rho\right)\pa_i v=\bar\rho_m\pa_i v_m+\frac43\bar\rho_\ga\pa_i v_\ga\,,\quad
        \end{split}
    \eneq
where~$\rho$,~$p$, and~$v$ are the energy density, pressure, and velocity potential of the fluid respectively. Individual fluid quantities are labeled by~$_m$ to denote pressure-less matter and by~$_\ga$ to denote radiation. Unlike the velocity potential, the energy density and pressure have non-vanishing background values~$\bar\rho$~and~$\bar p$, with the individual fluid contributions further satisfying the equation of state~$\bar p_m=0$ for non-relativistic matter and~$\bar p_\ga=\frac13\bar\rho_\ga$ for radiation. 

The metric and the energy-momentum tensor in Eqs.~($\ref{metric}$) and~($\ref{energymomentum}$) are the key ingredients needed to derive the theoretical expressions of the cosmological observables probed by galaxy surveys and CMB experiments. By solving the null geodesic equations for light propagation, and equipping the spacetime with a tetrad field to model observations \cite{Tetrad}, we can derive the linear-order relativistic expressions of cosmological probes such as the luminosity distance~$\de D_L$ \cite{LF}, galaxy clustering~$\de_g$ \cite{GC}, and the CMB temperature anisotropies~$\Theta$ \cite{CMB}. The gauge-invariant expressions derived in this way are independent of the underlying theory of gravity and take into account all the relativistic effects \cite{gi} at linear order in relativistic perturbation theory.

These relativistic effects include, for example, the gravitational redshift, which introduces perturbations proportional to the gravitational potential into the theoretical descriptions of the probes. Such perturbations can, in principle, dominate the statistics of the probes, even to the point of causing divergences when infinite wavelength perturbations are considered. To illustrate this concept, consider the variance in a given survey of one of the probes mentioned above, e.g. galaxy clustering:
    \bear\label{variance}
        \langle\de^2_g(x)\rangle=\int_0^\infty \frac{dk}k~\rvert\mathcal{T}_{\de_g}(\eta,k)\rvert^2\Delta^2_\RR(k)~,
    \enar
where~$\mathcal{T}_{\de_g}$ is the transfer function and~$\Delta^2_\RR=A_{\text{S}}(k/k_p)^{n_{\text{S}}-1}$ is the nearly scale-invariant primordial power spectrum with scalar spectral index~$n_{\text{S}}\simeq1$ and amplitude~$A_{\text{S}}\simeq10^{-9}$ at a given pivot scale~$k_p$.  The presence of the gravitational potential in the cosmological probe implies an enhanced sensitivity to long wavelength fluctuations: if the spectrum of curvature perturbations remains scale invariant with a small red tilt on arbitrarily large length scales, then the integral is dominated by the contributions at very low Fourier modes~$k$ and may become divergent as on large scales the transfer function of the gravitational potential is constant in~$k$. Indeed, such an infrared divergence signals the breakdown of our theoretical description.
 
 In  \cite{Ghazal1, Comeau} it was already noted that in computing the effects of long wavelength fluctuations, great care must be taken as to whether these effects are visible in local physical observables.  Thus, we must make sure that an infrared divergence such as the one pointed out above is not an artifact of the theoretical description.  In a previous work \cite{MY}, we showed that if the long wavelength fluctuations in the theoretical expression of the probe obey specific conditions, then precise cancellations occur such that any observable statistics as in Eq.~($\ref{variance}$) are finite.  On the other hand, if these conditions are not fulfilled, the statistics retain an enhanced infrared sensitivity and potentially diverge. In the remainder of this section, we review the main results of \cite{MY} in light of what will be investigated in this paper.

Given the theoretical expressions of the probes~$\de D_L$,~$\de_g$, and~$\Theta$ obtained as explained above (see, e.g., \cite{MY} for the specific equations), we can extract the contribution of fluctuations with wavelengths much larger than the scale of the probe~$\rbar$ by averaging in position space on a comoving scale~$R\gg\rbar$. After averaging, a generic perturbation~$f(x)$ affects the probe via its long wavelength contribution~$f_L(x)$, which satisfies the following expansion around an arbitrary origin which we can take to be the observer's position:
   \bear\label{01}
        f_L(x)&=&f_L(\eta)+x^i\pa_i f_L(\eta)+\OO(\bm x^2)
        \nnn
        &\equiv&f_0(\eta)+\rbar f_1(\eta,\bm{\hat{n}})+\OO(\bm x^2)~,
    \enar
where we expressed the comoving coordinates in terms of the comoving distance and the observed angles $x^i=\rbar\hat{n}^i$. Note that higher order terms in the expansion are suppressed by powers of~$\rbar/R$.

The coefficients~$f_0$ and~$f_1$ are those that most control the infrared sensitivity, as is evident from their Fourier space counterparts~$\Tilde{f}$, where we have expanded the Fourier phase for~$k\rbar\ll1$:
    \bear\label{01F}
        f_0(\eta)&=&\int \frac{d^3k}{(2\pi)^3}~\Tilde{f}(\eta,\bm k)~,
        \nnn
        f_1(\eta,\bm{\hat n})&=&\int \frac{d^3k}{(2\pi)^3}~ik_\para \Tilde{f}(\eta,\bm k)~.
    \enar
where $k_\para$ is the projection of the Fourier mode~$\bm{k}$ along the line-of-sight angular direction~$\bm{\hat{n}}$.

The general conditions governing the infrared sensitivity in \cite{MY} are conveniently expressed in terms of the comoving gauge and uniform density gauge curvature perturbations~$\RR$ and~$\zeta$, which for an individual fluid~$I$ are defined as
    \bear\label{defcurv}
        \RR_I\equiv\varphi-\HH v_I\,,\qquad\quad \zeta_I\equiv\varphi-\HH\frac{\de\rho_I}{\bar\rho'_I}\,,
    \enar
where~$I=m,\ga$ and~$\HH=\frac{a'}{a}$ is the conformal Hubble rate.
The corresponding total curvature perturbations are obtained by replacing~$v_I$ with the total velocity potential~$v$, and~$\de\rho_I$ with the total density~$\de\rho$. Note that the total energy density is the linear sum of the individual densities, while the total velocity potential is the weighted sum, as can be seen from the energy-momentum tensor in Eq.~($\ref{energymomentum}$). The main result of our previous work \cite{MY} is expressed by the following conditions,\footnote{Here we expressed the conditions for infrared insensitivity in a slightly different, but entirely equivalent way to that presented in \cite{MY}. }which we call conditions for infrared insensitivity:
    \begin{gather}
        \left(\frac{\de\rho_m}{\bar\rho'_m}-\frac{\de\rho_\ga}{\bar\rho'_\ga}\right)_0=0\,,\qquad v_{m,1}(\eta_{\bobs})=v_{\ga,1}(\eta_{\bobs})\,,
        \nnn\label{conditions}
        \left(\RR_m-\zeta_m\right)_{0,1}=0\,,
    \end{gather}
where~$\eta_{\bobs}=\int_0^\infty~dz/H$ is the observer's conformal time evaluated in a background universe.
If these relations are satisfied, then the first two coefficients in the long wavelength expansion of Eq.~($\ref{01}$) give no contribution to the cosmological probes:
    \bear
        (\de D_L)_{0,1}=0\,,\qquad (\de_g)_{0,1}=0\,,\qquad \Theta_{0,1}=0\,,
    \enar
so that the terms that potentially dominate the observable statistics exactly cancel out.

If perturbations are generated according to the simplest single-field inflationary mechanism, adiabatic initial conditions between matter and radiation are maintained on large scales, so that the equations in the first line of~($\ref{conditions}$) are satisfied. The equation in the second line of~($\ref{conditions}$), on the other hand, is a consequence of the assumed theory of gravity and the resulting equations of motion. For Einstein equations of general relativity, it is known that $\RR=\zeta$ is the only admissible solution up to~$\OO(k^2)$ differences, hence cosmological probes in the minimal $\Lambda$CDM model with adiabatic fluctuations satisfy the conditions for infrared insensitivity. For modified gravity models such as Horndeski theory, it has been shown \cite{MY2} that it is still possible to achieve the infrared insensitivity of the cosmological probes, but a strict tuning of the initial conditions is required to provide the necessary generalized adiabatic condition involving the additional scalar degree of freedom.

In this paper, we assume general relativity as the theory of gravity but consider a model that explicitly violates the adiabatic condition on large scales, potentially breaking all the conditions for infrared insensitivity.

        \section{FLUCTUATIONS IN A MODEL WITH INITIAL AXION ISOCURVATURE FLUCTUATIONS}
\label{sec3}

Most of the previous work on the infrared sensitivity of cosmological probes to super-Hubble fluctuations was done in the context of adiabatic fluctuations. Matter, however, consists of a large number of fields, and hence it is easily possible that the spectrum of fluctuations will have an isocurvature component. This can occur, for example, if we consider the effects of an axion field in cosmology. Axions \cite{QCDaxion} are well motivated from the point of view of particle physics since the introduction of an axion can solve the strong CP problem of particle physics \cite{PQ}. Axions are, in fact, a promising candidate for dark matter \cite{AxionDM}. The QCD axion is a pseudoscalar field which is massless at high temperatures but acquires a mass through non-perturbative effects at the QCD scale. As was discussed in \cite{Minos}, axions will at that time generically lead to axion-isocurvature fluctuations on super-Hubble scales which then seed growing curvature perturbations.

Axions also arise generically in low-energy effective field theories stemming from superstring theory (see e.g. the review in \cite{Witten}). Hence, our analysis is applicable in a broader range of settings, not just in the case of the QCD axion.

We will consider matter fields to consist of two components, radiation and the axion field. It is easy to include a separate matter component (be it baryonic or non-axionic cold dark matter), but doing this would only clutter the notation. There are two key energy scales associated with the axion, first the Peccei-Quinn symmetry breaking scale~$f_\chi$ (typically many orders of magnitude larger than the QCD scale) when the axion field materializes, and second the QCD scale when the axion acquires a mass. The fluctuations in the axion field depend on the cosmological scenario before the QCD scale. If there was a period of inflation after~$f_\chi$ (i.e. if~$f_\chi$ is higher than the energy scale of inflation), then the axion field within our present Hubble patch will take on (modulo quantum fluctuations) a fixed value. On the other hand, if~$f_\chi$ is smaller than the energy scale of inflation (or if the primordial universe did not involve a phase of inflation), then we expect the axion field to take on fluctuations of magnitude~$f_\chi$ within the current Hubble patch. In either case, we expect that the axion fluctuations will not be correlated with the initial curvature fluctuations, and hence there will be an isocurvature —or, more generally, entropy—component to the fluctuation spectrum.

In the following we will study the evolution of fluctuations during the various periods of the primordial universe. To be specific, we will consider the case of pre-inflationary Peccei-Quinn symmetry breaking. At the end we will comment on the changes if there was no inflation (or if Peccei-Quinn symmetry breaking happened after inflation).

        \subsection{During Inflation}

Let~$\phi(x)$ be the canonically normalized inflaton scalar field, and~$\chi(x)$ be the axion field (also with canonical normalization). We assume an early period of accelerated expansion driven by the potential energy of the inflaton field as in the single-field slow-roll inflation paradigm \cite{Guth}.\footnote{The reader should be aware of the challenges \cite{swamp, TCC} which this paradigm is facing from the point of view of fundamental theory.} During inflation, the axion is a massless spectator field, and the energy-momentum tensor of the matter fields is dominated by the inflaton:
    \bear
	T_{\mu\nu}=T^{(\phi)}_{\mu\nu}+T^{(\chi)}_{\mu\nu}\approx T^{(\phi)}_{\mu\nu}\,.
    \enar
The axion couplings are suppressed by~$f^{-1}_\chi$, hence there is effectively no interaction between the inflaton and the axion field, and the energy-momentum tensors are separately covariantly conserved: 
    \bear\label{KG}
	\nabla_\nu T_{(\phi)}^{\mu\nu}=0\,,\qquad \nabla_\nu T_{(\chi)}^{\mu\nu}=0\,.
    \enar
The Klein-Gordon equations resulting from the equations above have the key difference that the inflaton potential~$V(\phi)$ is non-trivial, while $V(\chi)=0$ since the axion is massless and interactions are suppressed.

For an FLRW universe, Eq.~($\ref{KG}$) reduces to
    \bear
	\Bar{\phi}''+2\HH\Bar{\phi}'+a^2V_{,\phi}=0\,,\Dquad\Bar{\chi}''+2\HH\Bar{\chi}'=0\,,~~
    \enar
where~$V_{,\phi}=\frac{dV}{d\phi}$ is the derivative of the inflaton potential and the overbars denote the time-dependent background values of the inflaton and axion field. The equation for the axion can be readily integrated yielding the background solution
    \bear\label{dotchi}
	\bar{\chi}'\propto a^{-2}\,,
    \enar
which, due to the accelerated expansion of the universe, implies that~$\Bar{\chi}$ rapidly relaxes to a constant characterizing our inflationary patch. 

During inflation, quantum fluctuations of the inflaton~$\de\phi$ are coupled to metric fluctuations~$\de g_{\mu\nu}$ in a gauge-dependent way. The standard procedure for dealing with scalar fluctuations during inflation is to fix the comoving gauge~$v_\phi\equiv 0\equiv\ga$, where~$\ga$ is a perturbation in the spatial components of the metric, and~$v_\phi$ is the velocity potential associated to the inflaton fluctuations:
    \bear
	v_{\phi}=\frac{\de\phi}{\Bar{\phi}'}\,.
    \enar
In the comoving gauge, the inflaton does not fluctuate and all the information is retained by the comoving gauge curvature perturbation~$\RR$, which is conserved on super-Hubble scales. The dimensionless power spectrum follows from the standard slow-roll inflation analysis (see e.g. \cite{MFB, RHBfluctsrev, Baumann} for reviews):
    \bear
	\Delta^2_\RR&=&\frac{H^2}{8\pi^2M^2_{pl}\epsilon_{sr}}\bigg\rvert_{k=aH}\equiv A_{\text{S}}\left(\frac{k}{k_p}\right)^{n_{\text{S}}-1}\,,
        \nnn
        n_{\text{S}}-1&=&-4\epsilon_{sr}+2\eta_{sr}\,,
    \enar
where ~$\epsilon_{sr}$ and~$\eta_{sr}$ are slow-roll parameters.

We pointed out that the axion field is a massless spectator during inflation. Nevertheless, it does acquire quantum fluctuations~$\de\chi$ with a nearly scale-invariant power spectrum. To motivate this statement we first derive the linear-order Klein-Gordon equation for the axion fluctuation
    \bear\label{dechi}
	\de\chi''+2\HH\de\chi'+k^2\de\chi=\Bar{\chi}'\left( \al'-3\varphi'+k^2\be+k^2\ga' \right)\,,\quad
    \enar
and compare it with the well-known equation of motion for tensor polarizations~$h$ during inflation
    \bear
	h''+2\HH h'+k^2h=0\,.
    \enar
At first sight, the two equations differ, however, since~$\Bar{\chi}'$ quickly approaches zero during inflation [Eq.~($\ref{dotchi}$)] we can neglect the source term of~$\de\chi$ and the two equations coincide.  It follows that the quantization procedure of the axion fluctuation is analogous to that of tensor fluctuations, modulo a factor of~$M^2_{pl}/2$ times the number of independent polarizations to account for the different coupling to gravity. The power spectrum of the axion fluctuations follows then the statistics:\footnote{Assuming the standard consistency relation between the tensor tilt and the amplitude of the spectrum of curvature fluctuations which holds in simple single-field models of inflation.}
    \bear\label{pschi}
        \Delta^2_{\de\chi} \, = \, \frac{H^2}{4 \pi^2} \, = \, \frac{M^2_{pl}}{4}\Delta^2_h\,,
    \enar
where the dimensionless power spectrum of the tensor polarizations reads
    \bear\label{psh}
        \Delta^2_h&=&\frac{H^2}{\pi^2M^2_{pl}}\bigg\rvert_{k=aH}\equiv\frac12 A_{\text{T}}\left(\frac{k}{k_p}\right)^{n_{\text{T}}(k)}\,,
        \nnn
        n_{\text{T}}&=&-2\epsilon_{sr}\,.
    \enar
Note that this relation only applies if $H < f_{\chi}$ since the fluctuations of $\chi$ cannot be larger than its range.

Unlike tensor fluctuations,~$\de\chi$ is in principle a gauge-dependent quantity, and therefore the power spectrum~$\Delta^2_{\de\chi}$ should depend on the chosen gauge. However, such dependence is irrelevant because under a coordinate transformation $x^\mu\to\tilde x^\mu=x^\mu+\xi^\mu$, the axion fluctuations~$\de\chi$ gauge transform as
    \bear
	\widetilde{\de\chi}=\de\chi-\xi^0\Bar{\chi}'\,,
    \enar
which again from the vanishing of~$\Bar{\chi}'$ implies that~$\de\chi$ is effectively gauge-invariant. This statement holds as long as the axion is massless.

Finally, we emphasize that the axion field fluctuations are generated independently of the inflaton field fluctuations and do not perturb the total energy density during inflation. Thus, entropy perturbations are induced at a later time, when the inflaton has decayed and the axion contribution to the energy-momentum tensor becomes important.

        \subsection{End of Inflation}

Inflation has to connect with the Standard Cosmology radiation phase. This is typically modeled by the inflaton decaying into radiation (see e.g. \cite{RHrevs} for reviews of the reheating dynamics). Matter can also be produced in this process. In particular, if we consider a setup in which the axion field does not provide sufficient energy to explain all of the dark matter, there would be a cold dark matter component which is also generated during reheating (denoted by $c$). Since it is reasonable to assume that radiation and cold dark matter are produced in proportion, we assume their fluctuations to be adiabatic on super-Hubble scales,
    \bear\label{adiabatic}
	\frac{\de_\ga}4=\frac{\de_c}3\,,\Dquad v_\ga=v_c~, 
    \enar
where~$\de_I=\frac{\de\rho_I}{\Bar{\rho}_I}$ is the density contrast of species~$I=\ga, c$. The equation above can be equivalently written in terms of individual curvature perturbations as
    \bear
        \zeta_\ga=\zeta_c\,,\Dquad \RR_\ga=\RR_c\,.
    \enar
Since the $c$ component plays no role in the generation of the entropy perturbations, we will not consider it further.

As long as the axion has no potential, the fluctuations~$\de\chi$ carry (to linear order in the amplitude of the fluctuations) no energy, and the energy density of the axion is negligible. Moreover, from the Einstein equations, it follows that the two gauge-invariant curvature perturbations~$\RR$ and~$\zeta$ are equal and conserved on super-Hubble scales \cite{Bardeen}
    \bear
	\RR=\RR_\ga=\zeta_\ga=\zeta\,, \Dquad \RR'=0=\zeta'\,.
    \enar
This description is valid until the QCD scale when the axion acquires a mass. Note that had we been tracking the~$c$ component, then we would have equalities among the individual~$\RR_I$ and~$\zeta_I$ as a consequence of adiabaticity.

        \subsection{At the QCD Scale}
\label{sec3C}

We assume that the axion suddenly acquires a potential at the QCD scale, corresponding to the temperature $T_{\text{QCD}}\approx 150\, \up{MeV}$
    \bear
	V(\chi)\approx m^2_\chi f^2_\chi\left[ 1-\cos\left(\frac{\chi}{f_\chi}\right) \right] \,.
    \enar
In the context of pre-inflationary Peccei-Quinn symmetry breaking, the axion field will be (modulo quantum fluctuations) coherent across the current Hubble patch, but its value will typically not align with the value of~$\chi=0$ which minimizes the potential. Hence,  $\chi$ will begin a period of oscillations about the vacuum state, yielding an equation of state whose time average corresponds to non-relativistic matter. In practice, from the QCD scale onwards, we treat the axion field as a pressure-less fluid with equation of state~$w_\chi=0$.

At the QCD scale, the universe does not undergo a phase transition, as radiation remains the dominant component. This implies that the total energy-momentum tensor and the metric components are continuous over the QCD scale, because~$\eta_{\text{QCD}}$ is just like any other time in the radiation-dominated era as far as total quantities are concerned:
    \bear\label{contot}
	g_{\mu\nu}\big\rvert_-=g_{\mu\nu}\big\rvert_+\,,\Dquad T_{\mu\nu}\big\rvert_-=T_{\mu\nu}\big\rvert_+\,, 
    \enar
where~$\rvert_-$ denotes a quantity evaluated at a time right before the QCD scale, and~$\rvert_+$ right after. 

While total quantities are continuous, the assumption of instantaneous energy transfer from radiation to axion at the QCD scale implies that the individual energy densities are discontinuous. At the background level, the energy transfer is negligible, while at linear order in perturbation theory Eq.~($\ref{contot}$) implies
    \bear\label{conti}
	\de\rho_\ga\big\rvert_-&=&\de\rho_\ga+\de\rho_\chi\big\rvert_+\,,
    \enar
showing that the initial radiation energy density is split into radiation and axion energy density when the axion becomes massive.

The energy transfer that we are considering happens at a fixed temperature, or in other words on a time slice with fixed radiation energy density.  The uniform density gauge condition~$\de\rho_\ga\rvert_-\equiv0$ identifies exactly the hypersurface of interest so it is the most convenient choice to perform computations. Using Eq.~($\ref{conti}$) we can express the radiation curvature perturbation~$\zeta_\ga$ right after the energy transfer occurred:
    \bear\label{part}
	\zeta_\ga=\zeta_--\frac{\de\rho_\chi}{4\bar\rho_\ga}\,,
    \enar
where we used that in the uniform density gauge~$\varphi=\zeta_-$. Moreover, the Einstein equations in the uniform density gauge set the total velocity potential~$v$ to zero on super-Hubble scale, which implies $4\bar\rho_\ga v_\ga+3\bar\rho_\chi v_\chi=0$. Combining the previous equations we derive the relative entropy perturbation and the relative velocity among axion and radiation:  
    \bear\label{rel}
	\zeta_{\chi\ga}&:=&\zeta_\chi-\zeta_\ga=\left( 1+\frac{3\bar\rho_\chi}{4\bar\rho_\ga} \right)\frac{\de_\chi}3\,,
	\nnn
	v_{\chi\ga}&:=&v_\chi-v_\ga=\left( 1+\frac{3\bar\rho_\chi}{4\bar\rho_\ga} \right)v_\chi\,,
    \enar
where the large round brackets are of order one.

Treating the axion field as a pressure-less fluid implies that the time average of the axion energy density is related to the time average of the axion potential as
    \bear
        \Bar{\rho}_\chi=2 V(\Bar{\chi})\,,\Dquad \de\rho_\chi=2\de\chi V_{,\chi}\,,
    \enar
with~$V \sim \chi^2$. Taking into account the scaling~$\Bar{\rho}_\chi\propto a^{-3}$ and the definition of the axion fluid velocity potential~$v_\chi=\frac{\de\chi}{\Bar{\chi}'}$ we recast Eq.~($\ref{rel}$) in the following form
    \bear\label{nad}
        \zeta_{\chi\ga}=-\HH v_{\chi\ga}\,,
    \enar
which quantifies the amount of non-adiabaticity set at the QCD scale in a gauge-independent way.

Notice that the right-hand-side of the equation above is nothing but the difference~$\RR_\chi-\RR_\ga$ of the individual comoving curvature perturbations. On super-Hubble scales, the equality~$\RR=\zeta$ given by the Einstein equation implies the following conditions at the QCD scale
    \bear\label{init}
        \RR_\chi=\zeta_\chi\,,\Dquad\RR_\ga=\zeta_\ga\,.
    \enar
Eqs.~($\ref{nad}$) and~($\ref{init}$) will be used as initial conditions for the evolution of the curvature perturbation.

        \subsection{After the QCD transition}

After the sudden interaction, we assume that axion and radiation evolve independently. The separate covariant conservation of the energy-momentum tensors implies the super-Hubble conservation of the individual~$\zeta_I$:
    \bear
	\zeta'_\ga=0\,, \Dquad \zeta'_\chi=0\,.
    \enar
As a consequence, the amount of non-adiabaticity~$\zeta_{\chi\ga}$ set at the QCD scale [Eq.~($\ref{nad}$)] remains constant on super-Hubble scales throughout the evolution of the universe.

The total curvature perturbations~$\RR$ and~$\zeta$ are still equal on super-Hubble scales thanks to the Einstein equations, however, they are no longer constant in time due to the presence of non-adiabatic fluctuations. The covariant conservation of the total energy-momentum tensor in the uniform density gauge implies, on super-Hubble scales
    \bear\label{Z'}
	\zeta'=-\frac{\HH}{\Bar{p}+\Bar{\rho}}\de p\,,
    \enar
which shows that~$\zeta$ is constant only in the case of purely adiabatic fluctuations since~$\de p=c^2_s\de\rho$ is zero in the uniform density gauge.

In our model, the pressure fluctuation is given by the radiation pressure $\de p_\ga=\frac13\de\rho_\ga$, which can be replaced in favor of~$\zeta_{\chi\ga}$ manipulating the uniform density gauge condition $\de\rho_\chi+\de\rho_\ga=0$. After straightforward calculations we recast Eq.~($\ref{Z'}$) as
   \bear\label{R'}
	\zeta'=\frac{12\HH\bar\rho_\ga\bar\rho_\chi}{(4\bar\rho_\ga+3\bar\rho_\chi)^2}\zeta_{\chi\ga}\,.
    \enar
Note that the whole time dependence is contained in the prefactor of~$\zeta_{\chi\ga}$, since the entropy mode is conserved on super-Hubble scales.
To integrate Eq.~($\ref{R'}$) we consider the background solutions of a radiation plus dust universe, where the role of dust is played by the axion
    \bear\label{radust}
	a=a_{eq}\left(2\xi+\xi^2\right)\,,\quad
	\bar\rho_\chi=\frac{\bar\rho_{eq}}2\frac{a^3_{eq}}{a^3}\,,\quad
	\bar\rho_\ga=\frac{\bar\rho_{eq}}2\frac{a^4_{eq}}{a^4}\,,\quad~~
    \enar
where we defined the dimensionless time variable~$\xi$ normalizing by the conformal time~$\eta_{eq}$ at matter-radiation equality
    \bear\label{xi}
	\xi\equiv\frac\eta{\Tilde{\eta}_{eq}}\,\Dquad \Tilde{\eta}_{eq}\equiv(1+\sqrt{2})\eta_{eq}\,.
    \enar
Integrating from the QCD scale~$\xi_{\text{QCD}}$ up to a generic time~$\xi$ we derive the time evolution of the curvature perturbation 
    \bear\label{R}
	\zeta=\zeta_\chi-\frac{4}{1+3(\xi+1)^2}\zeta_{\chi\ga}\,,
    \enar
where we chose the integration constant $\zeta(\xi_{\text{QCD}})=\zeta_\ga$ since the QCD scale is deep in the radiation-dominated era: $\xi_{\text{QCD}}\approx\frac{T_{\text{eq}}}{T_{\text{QCD}}}\approx10^{-9}$.

So far we have used the temporal component of the covariant conservation of the energy-momentum tensors to derive that the uniform density gauge curvature perturbations~$\zeta_\chi$ and~$\zeta_\ga$ are constant in time on super-Hubble scales. We now show that the spatial components of the covariant conservation impose that~$\RR_\chi$ and~$\RR_\ga$ can \textit{not} be constant if the adiabatic condition is violated. In an arbitrary gauge,~$\nabla_\mu T^{\mu i}_{(m)}=0=\nabla_\mu T^{\mu i}_{(\ga)}$ implies the following equations for the individual velocity potentials on super-Hubble scales:
    \bear
	v'_\ga-\frac{\de_\ga}4-\al=0\,,\qquad~ v'_\chi+\HH v_\chi-\al=0\,,
    \enar
which can be combined to derive an equation for the relative velocity potential
    \bear
	v'_{\chi\ga}+\HH v_{\chi\ga}=\RR_\ga-\zeta_\ga ~.
    \enar
We then rearrange the right-hand-side using a manipulation of the definition of~$\RR$ as a mixture of dust and radiation:
    \bear\label{RgTemp}
	\RR=\frac{3\bar\rho_\chi\RR_\chi+4\bar\rho_\ga\RR_\ga}{3\bar\rho_\chi+4\bar\rho_\ga}=\RR_\ga-\frac{3\HH\bar\rho_\chi}{4\bar\rho_\ga+3\bar\rho_\chi}v_{\chi\ga}\,.
    \enar
and substitute Eq.~($\ref{R}$) for~$\RR$ (recall $\RR=\zeta$ on super-Hubble scales) to obtain a differential equation for the relative velocity potential
    \bear
        v'_{\chi\ga}+\frac{4\HH\bar\rho_\ga}{3\bar\rho_\chi+4\bar\rho_\ga}v_{\chi\ga}-\frac{3\xi(2+\xi)}{1+3(\xi+1)^2}\zeta_{\chi\ga}=0\,.
    \enar
Integrating the equation above using the expressions for a radiation plus dust universe in Eq.~($\ref{radust}$) we obtain
	\bear
	\frac{v_{\chi\ga}}{\tilde\eta_{eq}}=\frac{8+12\xi+6\xi^2}{3\xi(2+\xi)}\left(C+\zeta_{\chi\ga}\right)+\frac{\xi^2}{2+\xi}\zeta_{\chi\ga}\,.
	\enar
We fix the integration constant~$C(\bm x)$ by matching the initial condition in Eq.~($\ref{nad}$)
	\bear
	C=-\zeta_{\chi\ga}\left(1+\frac34\xi^2_{\text{QCD}}\right)\,,
	\enar
neglecting $\OO(\xi_{\text{QCD}}^3)$. This choice leads to the expression
	\bear\label{vrel}
	\frac{v_{\chi\ga}}{\tilde\eta_{eq}}=\frac{2\xi^3-\xi^2_{\text{QCD}}\left(4+6\xi+3\xi^2\right)}{2\xi(2+\xi)}\zeta_{\chi\ga}\,,
	\enar
which indeed satisfies the initial condition in Eq.~($\ref{nad}$) since~$\HH^{-1}_{\text{QCD}}={\xi_{\text{QCD}}\tilde\eta_{eq}}$.
 
Given the relative velocity we can now compute~$\RR_\ga$ as a function of time via Eqs.~($\ref{RgTemp}$) and ($\ref{R}$):
	\bear\label{Rg}
	\RR_\ga=
        \zeta_\ga+3\frac{\xi^2-\xi^2_{\text{QCD}}(1+\xi)}{\xi(2+\xi)}\zeta_{\chi\ga}\,,
	\enar
and from the difference of the definitions:~$\RR_\ga-\RR_\chi=\HH v_{\chi\ga}$ we readily derive the expression for~$\RR_\chi$ as a function of time:
	\bear\label{Rm}
	\RR_\chi=\zeta_\chi-4\frac{\xi^2-\xi^2_{\text{QCD}}(1+\xi)}{\xi^2(2+\xi)^2}\zeta_{\chi\ga}\,.
	\enar
Indeed, Eqs.~($\ref{Rg}$) and ($\ref{Rm}$) are consistent with Eq.~($\ref{init}$) in the limit~$\xi\to\xi_{\text{QCD}}$.

We conclude that in the presence of an entropy fluctuation~$\zeta_{\chi\ga}$ the matter fluctuations~$\RR_\chi$ and~$\zeta_\chi$ do \textit{not} coincide on super-Hubble scales.  The difference between the two potentials scales as~$\xi^{-2}$ and thus is more important at early times than at later times. As we will see below, it follows that the signatures of the entropy fluctuations will be more pronounced in CMB observations (which probe the universe at the time of recombination which corresponds to a value of~$\xi\approx1$) than in large-scale structure surveys (which probe the universe at late times, for values~$\xi\gg1$).

In this section we have assumed that the Peccei-Quinn (PQ) symmetry breaking occurs before or during inflation such that the axion fluctuations are quantum fluctuations generated during inflation and squeezed on super-Hubble scales. If, on the other hand, the Peccei-Quinn symmetry breaking happens after inflation, then there will be no homogeneous axion field background across our Hubble patch. Rather, there will be fluctuations of magnitude~$\delta\chi\sim f_\chi$ on the comoving scale~$l_c$ which corresponds to the correlation length of the symmetry breaking phase transition, and which is bounded from above by the comoving Hubble length at the transition time. On scales larger than~$l_c$, the fluctuations are Poisson suppressed. Hence, in this situation we do not expect significant axion-induced entropy fluctuations on present-day cosmological scales.

The Ekpyrotic scenario \cite{Ekp} is an interesting alternative to cosmological inflation in terms of solving the usual problems of Standard Big Bang cosmology and providing an explanation for the currently observed fluctuations. The scenario involves a period of very slow contraction driven by matter with an equation of state~$p\gg\rho$ which can be obtained by postulating a scalar field with negative exponential potential. This is then followed by a non-singular cosmological bounce.\footnote{Such a bounce requires going beyond the usual effective field theory approach with matter satisfying the usual energy conditions.} If the cosmological bounce is obtained by the presence of an S-brane (a space-filling brane with vanishing energy density and negative pressure which appears once the energy density reaches a critical scale), then a scale-invariant spectrum of entropy fluctuations can be obtained at the beginning of the expanding period in the same way that a scale-invariant spectrum of gravitational waves is obtained \cite{Ziwei}. If we assume that the PQ symmetry is never restored (i.e. the energy scale of the bounce is lower than the PQ symmetry breaking scale), then the resulting spectrum of entropy fluctuations is like that computed in this section in the case of an inflationary scenario with PQ scale higher than the energy scale of inflation.

It would also be interesting to study the entropy fluctuations generated in an emergent cosmology such as String Gas Cosmology \cite{SGC} (see \cite{SGCrev} for a review) and Matrix Cosmology \cite{matrix} (see \cite{matrixrev} for a review) in which the universe begins in a quasi-static thermal state with thermal fluctuations and then undergoes a phase transition to the expanding phase of Standard Big Bang cosmology.

        \section{COSMOLOGICAL PROBES}

In the previous section we have studied how metric and matter fluctuations evolve when the presence of an axion field explicitly violates the adiabatic condition. We have shown that the violation of the adiabatic condition occurs at a precise scale, the QCD scale, and that the resulting entropy fluctuations~$\zeta_{\chi\ga}$ remain constant over time on large scales. Up to this point, the Hubble radius has been used to distinguish between large (super-Hubble) and small (sub-Hubble) scales, but for cosmological observables the natural separation of scales is given by the characteristic length of the probe, e.g., the comoving distance to the last scattering surface for the CMB anisotropies.

In Sec.~\ref{sec2} we have argued that fluctuations on scales much larger than the characteristic size of a given probe can be described by the long wavelength expansion in Eq.~($\ref{01}$). Importantly, the first two coefficients ($0$ and~$1$) of such expansion satisfy the same dynamics of super-Hubble fluctuations because second (spatial) derivatives vanish when acting on $f_0+\rbar f_1$. It follows that the infrared sensitivity of cosmological probes is governed precisely by the equations derived in Sec.~\ref{sec3}.

The cosmological observables considered in this paper probe the universe from the time of matter-radiation equality onward, so the dimensionless time variable~$\xi$ associated with the observed redshift is much larger than~$\xi_{\text{QCD}}$. As a result, Eqs.~($\ref{vrel}$)-($\ref{Rm}$) greatly simplify when applied to cosmological probes, in particular, the relative velocity potential and matter curvature fluctuations become:
    \bear\label{v01}
    (v_{\chi\ga})_{0,1}&=&\Tilde{\eta}_{eq}~\frac{\xi^2}{2+\xi}(\zeta_{\chi\ga})_{0,1}\,,
    \\
    \label{R01}
    (\RR_\chi-\zeta_\chi)_{0,1}&=&-\frac{4}{(2+\xi)^2}(\zeta_{\chi\ga})_{0,1}\,.
    \enar
The equations above manifestly \textit{violate} the conditions for infrared insensitivity in Eq.~($\ref{conditions}$), implying an enhanced infrared sensitivity of the cosmological probes in a model with axion fluctuations. In particular, if fluctuations were to extend to infinite wavelengths with an unchanged spectral index, then the presence of long entropy modes would cause infrared divergences in the statistics of all the cosmological probes.

For numerical evaluations, it is convenient to express the time variable~$\xi$ in terms of the observed redshift. From Eq.~($\ref{radust}$) for a radiation plus dust universe we obtain
    \beeq\label{xiz}
        2\xi_z+\xi^2_z=\frac{1+z_{eq}}{1+z}\,,
    \eneq
hence we expect~$\OO(\zeta_{\chi\ga})$ violation of the conditions for infrared insensitivity in the case of the CMB anisotropies, while~$\OO(z^{-1}_{eq}\zeta_{\chi\ga})$ violations for low redshift probes. In the following subsections we analyze in more detail the consequences of such violations.

        \subsection{IR sensitivity of low-redshift probes}

Despite the radiation density being negligible at low redshift, the violation of the adiabatic condition at the QCD scale is preserved until today and affects the infrared sensitivity of low redshift probes such as the luminosity distance and galaxy clustering.

As a concrete example, we compute the effects of the entropy mode on the luminosity distance fluctuation~$\de D_L$. The first two coefficients in the long wavelength expansions were derived in \cite{MY}:
    \bear
    \de D_L&=&\left[\RR_\chi\big\rvert_{\bobs}+\frac{\RR_\chi\big\rvert_z-\RR_\chi\big\rvert_{\bobs}}{\rbar_z\HH_z}-\frac{1}{\rbar_z}\int_0^{\rbar_z}d\rbar~\RR_\chi\right]_0
    \nnn
    &&
    +\frac{1}{\HH_z}\left[ \RR_\chi\big\rvert_z-\frac{1}{\rbar_z}\int_0^{\rbar_z}d\rbar~\RR_\chi\right]_1\,,
    \enar
where $\rvert_{\bobs}$ denotes evaluation at the observer position and $\rvert_z$ at the observed redshift. Substituting Eq.~($\ref{R01}$) and considering~$\xi_z\gg1$ we obtain
    \bear
    \de D_L\approx \left( \frac2{\xi_z\xi_{\bobs}}-\frac6{\xi^2_{\bobs}}\right)(\zeta_{\chi\ga})_0+\Tilde{\eta}_{eq}\left( \frac{2}{\xi_{\bobs}}-\frac{2}{\xi_z} \right)(\zeta_{\chi\ga})_1\,,\quad
    \enar
and for a redshift of approximately zero $\xi_z\approx\xi_{\bobs}$
    \bear
    \de D_L\approx -\frac{4}{z_{eq}}(\zeta_{\chi\ga})_0\approx -10^{-3}(\zeta_{\chi\ga})_0\,,
    \enar
where we used that the redshift corresponding to the matter-radiation equality is roughly~$z_{eq}\approx 3200$.

The back-reaction~$\Bar{D}_L\de D_L$ on the background luminosity distance~$\Bar{D}_L$ would provide a correction to the Hubble constant
    \bear
    H_0\to H_0(1-\de D_L)\approx H_0\left[1+10^{-3}(\zeta_{\chi\ga})_0\right]\,,
    \enar
definitely not enough to explain the Hubble tension, for which we would need $\de D_L\approx0.1$\,. Note that we have neglected the cosmological constant in our analysis, which is instead important at low redshifts. However, even accounting for the cosmological constant, we do not expect any alleviation of the Hubble tension.

As another example, let us consider the fluctuations $\delta_g$ in the galaxy number counts.  As derived in \cite{MY}, the leading infrared contributions~$\delta_{g, 0}$ and~$\delta_{g, 1}$ are given by (dropping the contributions from volume fluctuations)
\begin{eqnarray}
\delta_{g, 0} \, &=& \, 3b \left( \zeta_\chi - {\cal{R}}_\chi \big\rvert_z\right)+e_z\left( {\cal{R}}_\chi \big\rvert_z-{\cal{R}}_\chi \big\rvert_{\bobs} \right) \\
\delta_{g, 1} \, &=& \, 3b \left( \zeta_\chi - {\cal{R}}_\chi \big\rvert_z\right)+e_z 
\left( {\cal{R}}_\chi \big\rvert_z - \frac{1}{\rbar_z} \int_0^{\rbar_z} d\rbar~\RR_\chi \right) \nonumber
\end{eqnarray}
where $b$ is the linear bias, $e_z$ the evolution bias, and we suppressed the subscripts~$_0$ and~$_1$ in the right-hand-sides.

The above contributions are proportional to the late time difference between $\zeta_\chi$ and ${\cal{R}}_\chi$, and also to the change of ${\cal{R}}_\chi$ over the time interval that light takes to travel from the galaxies to us. From the above analysis it follows that the effects of non-adiabaticity on the late-time probe of galaxy clustering are suppressed by a factor of the order of $z^{-1}_{eq}$.

The axion model we are considering could affect another tension that has been extensively studied recently, the cosmic dipole tension \cite{dipole}. As proposed by Turner \cite{Turner}, we may be living in a "tilted universe" where our peculiar velocity is modulated by long wavelength entropy fluctuations. The effect of this modulation could explain why the cosmic dipole measured in galaxy surveys exceeds the prediction from CMB measurements by a factor of two. In our model, the effects of long wavelength entropy fluctuations on galaxy clustering have the same suppression as the effects on the luminosity distance fluctuations, so there is no noticeable back-reaction on the dipole of large-scale structures. However, the prediction from CMB measurements could be altered by the presence of long wavelength entropy modes \cite{domenech}, since the CMB dipole would carry an extra contribution obtained by evaluating the relative velocity in Eq.~($\ref{v01}$) at~$\xi=\xi_{\bobs}$:
    \bear
    v_{\text{CMB}}\to v_{\text{CMB}}-\eta_{\bobs}(\zeta_{\chi\ga})_1\,,
    \enar
where we assumed that the observer is comoving with axion dark matter, and considered the special case where the long wavelength entropy mode contribution is aligned with the CMB dipole. We conclude that since the relative velocity between the CMB fluid and the observer is not suppressed, a relaxation of the cosmic dipole tension may be possible with this mechanism.

        \subsection{IR sensitivity of the CMB}

The infrared sensitivity induced by entropy fluctuations to the CMB temperature anisotropies is much larger than that retained by low redshift probes. This is because the dimensionless time parameter~$\xi_*$ corresponding to the redshift of last scattering~$z_*$ is approximately one [Eq.~($\ref{xi}$)], which implies that the long wavelength potentials are of the same order as the entropy fluctuations. In detail, the infrared sensitivity of the CMB temperature anisotropies is controlled by \cite{MY}:
    \bear\label{thetazeta}
    \Theta=\bigg[\zeta_\ga-\RR_\chi\big\rvert_{\bobs}\bigg]_0+\rbar_*\left[\zeta_\ga-\frac{v_{\chi\ga}\big\rvert_*}{\rbar_*}-\frac1{\rbar_*}\int_0^{\rbar_*}d\rbar~\RR_\chi\right]_1 ,~~
    \enar
where $\rvert_*$ indicates evaluation at the redshift of last scattering~$z_*$.
Substituting~$\xi_*\approx1$ and~$\rbar_*\approx\Tilde{\eta}_{eq}\xi_{\bobs}$, together with Eqs.~($\ref{v01}$) and~($\ref{R01}$), we find
    \bear\label{theta}
    \Theta\approx-(\zeta_{\chi\ga})_0-\rbar_*(\zeta_{\chi\ga})_1\,,
    \enar
showing indeed that long wavelength entropy fluctuations directly control the infrared sensitivity of the CMB anisotropies, without any suppression factor.
Using the results of Sec.~\ref{sec3C} we recast Eq.~($\ref{theta}$) in terms of the axion field values at the QCD scale, right after the axion potential is generated 
    \bear\label{thetachi}
    \Theta\approx -\frac13\de_\chi\approx-\frac23\frac{\de\chi}{\Bar{\chi}}\,,
    \enar
where we suppressed the subscript~$_0$ and~$_1$.

Combining Eqs.~($\ref{dechi}$) and~($\ref{dotchi}$) we deduce that the ratio~$\de\chi_{0.1}/\Bar{\chi}$ is constant from inflation until right before the QCD scale. Neglecting the evolution of this ratio from the moment the axion potential is generated until it behaves like cold dark matter, we can relate the statistics of the long wavelength CMB fluctuations to the axion initial conditions combining Eqs.~($\ref{thetachi}$) and~($\ref{pschi}$):\footnote{If the axion does not make up the whole matter, we have to multiply the result by the ratio of the squares of today's density parameters~$(\Omega^2_\chi/\Omega^2_m)\big\rvert_{\bobs}$.}
    \bear\label{Dtheta}
    \Delta^2_\Theta\approx\frac49\frac{\Delta^2_{\de\chi}}{\Bar{\chi}^2}
     \approx \frac1{9\pi^2} \frac{H^2}{f^2_{\chi}} \approx \frac19\frac{M^2_{pl}}{f_\chi^2}\Delta^2_h\,,
    \enar
where we used that in pre-inflationary Peccei-Quinn symmetry breaking the homogeneous value of the axion field~$\Bar{\chi}$ has a random but unique value of the order of the axion decay constant~$f_\chi$.  Note that the above formula is only true if $H < f_{\chi}$ since the range of $\chi$ is $f_{\chi}$ and the fluctuation in $\chi$ obviously cannot be larger than the range. We also see that initial axion isocurvature fluctuations imprint on the CMB infrared modes the same statistics as primordial gravitational waves but \textit{amplified} by a factor $M^2_{pl}/f_\chi^2$, which is of the order of~$10^{16}$ for $f_\chi\approx10^{10}\text{GeV}$. However,  from the penultimate expression in (\ref{Dtheta}) we see that the spectrum of the fluctuation is bounded from above by a number of the order one.  For the above value of $f_{\chi}$, the final expression in (\ref{Dtheta}) is only valid if $H < f_{\chi}$, and for such a low value of $H$ the gravitational wave spectrum is suppressed. Thus, there is no inconsistency between the penultimate and final expressions in (\ref{Dtheta}).

Although the CMB dimensionless power spectrum in Eq.~($\ref{Dtheta}$) is determined by low Fourier modes with $k\rbar_*\ll1$, in principle, \textit{all} Fourier modes contribute to the observable statistics. As a consequence, the presence of entropy fluctuations on large scales will in any case affect the correlations of CMB anisotropies on observable scales, leading to an enhanced infrared sensitivity. Considering the variance $\langle\Theta^2(\bm{\hat{n}})\rangle$ as the simplest example of observable statistics, we find a large logarithmic contribution coming from the lowest Fourier modes (in the case of a spectrum of fluctuations induced by canonical slow-roll inflation):
    \bear\label{var}
\langle\Theta^2(\bm{\hat{n}})\rangle=\int_{k_{\text{min}}}^{k_{\text{max}}} \frac{dk}{k} ~\Delta^2_{\Theta}~\sim ~ A_{\text{T}} \frac{M^2_{pl}}{f^2_\chi}\ln \left(\frac{k_{\text{min}}}{k_*}\right)\,,
    \enar
where $k_*$ is the momentum scale (much smaller than the characteristic momentum scale of CMB observations) above which the infrared approximation ceases to be justified.

Assuming the standard paradigm of inflation, the above equation places constraints on the axion isocurvature model, as well as on any model that violates the conditions for infrared insensitivity.  If fluctuations were present on arbitrarily large scales, with an unchanged spectral tilt, then $k_{\text{min}} \rightarrow 0$,  and infrared divergences would invalidate the theoretical descriptions of any observable statistics.

However,  we know that inflation cannot be past eternal \cite{Borde}, and hence there is a minimal value of $k$ below which the spectrum changes.  Given a particular physical origin of isocurvature fluctuations and a particular value of $H$ during inflation, demanding that the perturbative analysis of infrared effects does not break down would provide an upper bound on the duration of inflation. In fact, it has been conjectured that in a consistent theory of quantum gravity, fluctuations on scales that are initially smaller than the Planck scale cannot be stretched to scales larger than the Hubble radius.  This conjecture is known as the Trans-Planckian Censorship Conjecture (TCC) \cite{TCC} (see, e.g., \cite{TCCrev} for a review) and implies a finite duration for inflation with the number of e-foldings~$N$ bounded by the condition $N<\ln(M_{pl}/H)$.  If, in addition, we demand a sufficient duration of inflation in order for inflation to be able to provide a causal explanation for the origin of structure, then the TCC requirement not only provides an infrared cutoff that renders the integral in Eq.~($\ref{var}$) finite, but it leads to an upper bound on the value of $H$ during inflation and hence on the tensor to scalar ratio $r \equiv A_{\text{T}}/ A_{\text{S}} $ ($r <10^{-30}$) that greatly reduces the effects of the violation of the conditions for infrared insensitivity.

        \section{DISCUSSION}

We have shown here that long wavelength entropy fluctuations on a scale~$R$ can affect cosmological observables on smaller scales~$\rbar$. These effects peak for observables
at the time of equal matter and radiation and decay as~$\eta^{-2}$ for later time observables. Hence, the effects of long wavelength fluctuations are more important for CMB fluctuations than for quantities related to late-time galaxy clustering.

The effect of infrared modes on the amplitude and spectral shape of the power spectrum of observables depends on the spectrum of the long wavelength fluctuations which we have not specified in this work. Since in particular CMB observables are affected, the presence of infrared isocurvature fluctuations could affect the estimation of the cosmological parameters from CMB observations such as Planck \cite{Planck}.

The effect we are discussing here is different from the well-known constraints on isocurvature modes from the current data. These constraints (see e.g.  \cite{isocurv}) are based on studying the effects of linearized isocurvature fluctuations on the same scales as those of observations. Such fluctuations will also arise in our scenario, and they will be of the type of cold dark matter isocurvature perturbations. A pure isocurvature mode of this type will change both the amplitude and the peak structure of the angular power spectrum of CMB anisotropies. As reviewed e.g. in \cite{Baumann}, for large angles the fractional temperature fluctuations~$\Delta T/T$ will be related to the relativistic curvature perturbation in the Newtonian gauge~$\Phi$ via~$\Delta T/T=2\Phi$ instead of~$\Delta T/T=\Phi/3$ (the result for adiabatic fluctuations). In addition, for a pure isocurvature mode the first acoustic peak is at a smaller angular scale than for adiabatic fluctuations. The current data constrain the amplitude of the isocurvature mode to be at most~$0.03$ of the amplitude of the adiabatic mode ($95\%$ confidence level \cite{isocurv}). The constraints on the slope of the isocurvature mode are weak. Hence, the isocurvature fluctuations on super-Hubble scales could be larger than that of the adiabatic mode without being in conflict with the linear perturbation theory constraints.

In light of our analysis it would be of great interest to study a concrete scenario in which primordial cold dark matter isocurvature fluctuations are generated and study the constraints on their amplitude which can be derived based on their effects on the spectrum of CMB anisotropies.

        \acknowledgments

The research of R.B. at McGill is supported in part by funds from NSERC and from the Canada Research Chair program.  M. M and J.Y. are supported in part by the Swiss National Science Foundation.


\end{document}